\documentclass{jpsj2}
\usepackage{graphicx}
\usepackage{dcolumn}
\usepackage{bm}
\begin{document}

\title{Composite excitation of Josephson phase and spin waves in Josephson junctions with ferromagnetic insulator}

\author{Shin-ichi HIKINO$^{1}$, Michiyasu MORI$^{2,4}$, Saburo TAKAHASHI$^{3,4}$, and Sadamichi MAEKAWA$^{2,4}$}
\inst{	
$^{1}$Computational Condensed Matter Physics Laboratory, RIKEN, Wako, Saitama 351-0198, Japan \\
$^{2}$Advanced Science Research Center, Japan Atomic Energy Agency, Tokai-mura, Ibaraki 319-1195, Japan \\
$^{3}$Institute for Materials Research, Tohoku University, Sendai 980-8577, Japan \\
$^{4}$CREST, Japan Science and Technology Agency,  Tokyo 100-0075, Japan
}

\date{\today}

\abst{
Coupling of Josephson-phase and spin-waves is theoretically studied in a superconductor/ferromagnetic insulator/superconductor (S/FI/S) junction. 
Electromagnetic (EM) field inside the junction and the Josephson current coupled with spin-waves in FI 
are calculated by combining Maxwell and Landau-Lifshitz-Gilbert equations.
In the S/FI/S junction, it is found that the current-voltage ($I$-$V$) characteristic shows {\it two} resonant peaks. 
Voltages at the resonant peaks are obtained as a function of the normal modes of EM field, 
which indicates a composite excitation of the EM field and spin-waves in the S/FI/S junction. 
We also examine another type of junction, in which a nonmagnetic insulator (I) is located at one of interfaces between S and FI. 
In such a S/I/FI/S junction, {\it three} resonant peaks appear in the $I$-$V$ curve, 
since the Josephson-phase couples to the EM field in the I layer. 
}

\kword{Fiske resonance, Josephson junction, superconductor, ferromagnetic insulator, spin-wave}

\maketitle

\section{Introduction}
The dc Josephson effect is characterized by the zero-voltage current through a thin insulating barrier sandwiched by two superconductors\cite{josephson}. 
This effect is a macroscopic quantum phenomenon involving phase coherence between two superconductors. 
When a finite voltage($V$)-drop appears in the junction, the difference in the phase of superconducting order parameter, i.e. 
{\it Josephson-phase} ($\theta$), oscillates 
with time according to $\partial \theta /\partial t = (2e/\hbar)V $, and the alternating current with frequency $(2e/\hbar)V$ flows in the junction. 
This ac Josephson effect is derived by the gauge invariance including $\theta$. 
The electromagnetic response dominated by $\theta$ shows a resonant behavior in the junction. 
When a dc magnetic field and the dc voltage are applied to the junction, 
the electromagnetic (EM) field is generated by spatially modulated ac Josephson current. 
In this case, the current-voltage ($I$-$V$) curve exhibits resonant peaks due to the resonance 
between the ac Josephson current and the EM field generated by the spatially modulated ac Josephson current itself. 
This is called {\it Fiske resonance} \cite{fiske, eck, coon, kulik, barone}. 

In recent years, a ferromagnetic Josephson junction composed of ferromagnetic metal (F) and 
superconductors (S's), i.e., S/F/S junction, has received much attention\cite{golubov_sfs1, buzdin_sfs1, ryazanov_sfs1, kontos_sfs1}. 
One of the interesting effects is the formation of $\pi$ state arising from the Zeeman splitting in F. 
In addition, the interaction between Cooper pairs and spin waves in F is also of importance in the transport properties in the S/F/S junction.
\cite{nussinov_sfs1, stakahashi_sfs1, bell_sfs1, houzet_sfs1, hikino_sfs1, konschelle_sfs1, yokoyama_sfn1, petkovic_sfs1, volkov_sfs1}. 
In a small junction, where the junction width is smaller than the 
Josephson penetration depth, the spin-wave excitation induced by the ac Josephson effect is observed\cite{petkovic_sfs1}. 
In the recent experiment in a S/F/S junction including a nonmagnetic insulator (I) in one of interfaces between S and F, 
it has been reported that the Fiske resonance has multiple structures that must be associated with the spin-wave excitation\cite{aprili-m2s}. 
Volkov $et$ $al$. have theoretically studied collective excitations in such a junction and 
reported an additional structure in the Fiske resonance induced by spin-waves\cite{volkov_sfs1}. 
In their theory, a nonmagnetic insulator is crucial to obtain the Fiske resonance coupled with spin-waves. 
On the other hand, another type of ferromagnetic Josephson junction composed of ferromagnetic insulator (FI) and two S's, i.e., S/FI/S junction, 
is also expected to show the similar multiple structures in the Fiske resonance. 
It has been reported that the dissipation effect in the S/FI/S junction is smaller than that in the S/F/S junction\cite{kawabata1,kawabata2}. 
Such a small dissipation in the S/FI/S junction is due to the small probability of quasi-particle excitation in the FI\cite{kawabata1,kawabata2}. 
The damping of spin-waves induced by the similar mechanism is also very small in the FI compared to the case in F\cite{hillebrands, kajiwara}. 
Therefore, the coupling between Josephson-phase and spin-waves can be observed more clearly in the S/FI/S junction. 

In this paper, we theoretically study a composite excitation of the Josephson-phase and spin-waves in the S/FI/S and S/I/FI/S junctions. 
First, we calculate the dynamics of Josephson-phase coupled with spin waves by using Maxwell and Landau-Lifshitz-Gilbert (LLG) equations. 
Second, we derive the dc Josephson current induced by the Fiske resonance. 
In the S/FI/S junction, two resonant peaks appear in a current-voltage curve for each mode of the EM field. 
These two resonant peaks may be associated with the direct coupling between spin-waves and the EM field inside the junction. 
We also discuss the Fiske resonance in the S/I/FI/S junction. 
The non-magnetic high resistive layer is sometimes important, since the magnetic dead layer exist in the
ferromagnetic insulator near the S/FI interface. 
Our results clearly show the difference between S/FI/S and S/I/FI/S junctions in the dispersion relations of the Fiske resonance.
In such a S/I/FI/S junction, we show that three resonant peaks appear in the $I$-$V$ curve for each mode of the EM field. 

The rest of this paper is organized as follows. 
In Sec. II, by combining the Maxwell and LLG equations in a S/FI/S junction, 
we formulate the dc Josephson current induced by the Fiske resonance. 
In Sec. III, the Fiske resonance is discussed in S/FI/S and S/I/FI/S junctions. 
Summary is given in Sec. IV. 

\section{Formulation of Fiske resonance in S/FI/S junction} 
\begin{figure}[t]
\begin{center}
\vspace{10mm}
\includegraphics[width=7cm]{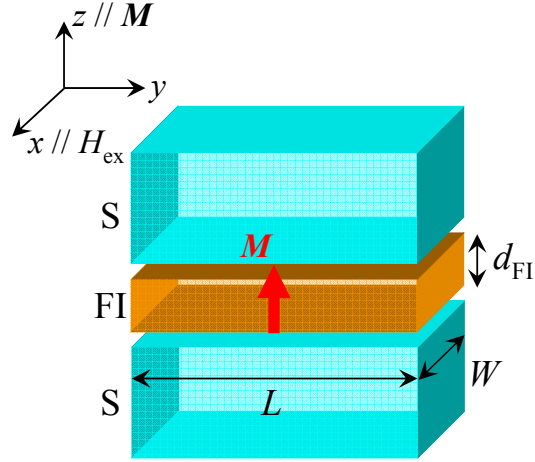}
\caption{ (Color online) Schematic figure of a S/FI/S junction with ferromagnetic insulator (FI) between two superconductors (S's). 
$d_{\rm FI}$ is the thickness of FI. 
$L$ and $W$ are the widths of the junction. 
$\bm M$ and $H_{\rm ex}$ are the magnetization in the FI and the applied dc magnetic field, respectively. 
}
\label{fjj-gm}
\end{center}
\end{figure}

The system considered is a Josephson junction with a FI sandwiched 
by two $s$-wave superconductors (S's) as shown in Fig.$~$\ref{fjj-gm}. 
The magnetization in the FI is parallel to the $z$-direction\cite{petkovic_sfs1}. 
A uniform dc magnetic field is applied in the $x$-direction. 
In the measurement of the Fiske resonance, the dc magnetic field is smaller than several tens of gauss. 
Therefore, we can neglect the in-plane magnetization induced by the applied dc magnetic field. 
Here, we consider that the ac electric and magnetic fields are in the $z$- and $x$-direction respectively, 
both of which are uniform in the $x$-direction. 
We consider the situation, in which the $z$-dependence of the electric and magnetic fields in the FI is negligible 
due to the very thin thickness of the FI ($d_{\rm FI}$). 
In the S regions, it is assumed that the magnetic field depends on $y$- and $z$-component. 
The current density has a nonzero $y$-component in the superconducting regions (Meissner current) and 
a nonzero $z$-component in the ferromagnetic region (quasi-particle and Josephson currents). 
Based on the above assumptions, the Maxwell equation in each region is given by 
\begin{eqnarray}
{\rm rot} \left[ E_{z}(y,t) {\bm e_{z}} \right] &=& 
-\frac{\partial }{\partial t} \left[\mu _{0} H_{x}(y,z,t) +M_{x}(y,t) \right] {\bm e}_{x},
\label{faraday} \\
{\rm rot} \left[ H_{x}(y,z,t) {\bm e}_{x} \right] &=&
J_{\rm M}^{y}(y,t){\bm e}_{y},
\label{ampere-sc} \\
{\rm rot} \left[ H_{x}(y,t) {\bm e}_{x} \right] - 
\frac{\partial }{\partial t} \left[ D_{z}(y,t) {\bm e}_{z} \right] &=&
J_{\rm J}^{z}(y,t){\bm e}_{z} + J_{\rm Q}^{z}(y,t){\bm e}_{z},
\label{ampere-fm} \\
M_{x}(y,t) &=& \int_{-\infty }^{\infty } dy'dt' 
\chi_{x} (y-y',t-t') H_{x}(y',t'), 
\label{mx} \\
J_{\rm J}^{z}(y,t) &=&J_{\rm c} \sin \theta(y,t)
\label{josephson-current}, \\
J_{\rm Q}^{z}(y,t) &=& \frac{1}{R_{\rm FI}} E_{z} (y,t) 
\label{qpc}.
\end{eqnarray}

Here, ${{\bm e}_{i}}$ ($i=x, y, z$) is a unit vector, 
$ E_{z}(y,t) $ is the electric field in the FI, 
$ H_{x}(y,z,t) $ and $H_{x}(y,t)$ are the magnetic fields in the S and the FI, respectively. 
The electrical flux density, $D_{z}(y,t)$, in FI is given by $D_{z}(y,t) = \epsilon_{\rm FI}E_{z}(y,t)$, 
where $\epsilon_{\rm FI}$ is the dielectric constant in FI. 
$ J_{\rm J}^{z}(y,t) $ and $J_{\rm c}$ are the Josephson current and the Josephson critical current densities, respectively. 
$ J_{\rm Q}^{z}(y,t) $ and $R_{\rm FI}$ are the quasi-particle current density and the resistivity of FI, respectively. 
The $x$-component of the magnetization, $ M_{x}(y,t) $, in the FI is given by Eq.$~$(\ref{mx}). 
The motion of magnetization is described by the Landau-Lifshitz-Gilbert (LLG) equation\cite{hillebrands}, 
\begin{eqnarray}
\frac{d{\bm M}}{dt} &=&
	-\gamma {\bm M}\times {\bm H_{\rm eff}}
	+\frac{\alpha }{M}
	\left[ 
	{\bm M}\times \frac{d{\bm M}}{dt}
	\right],
\label{mmeq}
\end{eqnarray}
where ${\bm M}$ is the magnetization of FI, $\gamma $ is the gyromagnetic ratio, and
$\alpha $ is the Gilbert damping. 
The effective field, to which ${\bm M}$ responds, is given by ${\bm H_{\rm eff}}$. 

By using Maxwell and LLG equations, we can obtain the voltage coupled with spin-waves 
(the detail of derivation for Eq.$~$(\ref{veq-sw}) is given in Appendix A.) as follows: 
\begin{eqnarray}
\frac{{\partial ^2 V \left( {y,t} \right)}}{{\partial y^2 }} &=& 
\frac{1}{{c_{\rm FI}^2 }}
\left[ {
{\frac{{\partial ^2 V \left( {y,t} \right)}}{{\partial t^2 }}}
+\frac{ d_{\rm FI }}{ d_{\rm FI}+2\lambda _{\rm L}} \frac{1}{\mu_{0}} 
\int_{-\infty }^{\infty } dy'dt' 
\chi (y-y',t-t') \frac{{\partial ^2 V \left( {y',t'} \right)}}{{\partial t^2 }} 
} \right. \nonumber \\
&+& 
\left. {
\Gamma_{\rm FI} \frac{\partial V(y,t)}{\partial t}
+ \Gamma_{\rm FI} \frac{d_{\rm FI}}{d_{\rm FI}+2\lambda _{\rm L}} \frac{1}{\mu_{0}}
\int_{-\infty }^{\infty } dy'dt' 
\chi (y-y',t-t') \frac{{\partial V \left( {y',t'} \right)}}{{\partial t }} 
 } \right] \nonumber \\
&+&
\frac{1}{\lambda_{\rm J}^{2}J_{\rm c}} J_{\rm J}^{z}(y,t)
+ \frac{1}{\lambda_{\rm J}^{2}J_{\rm c}} \frac{d_{\rm FI}}{d_{\rm FI}+2\lambda _{\rm L}} \frac{1}{\mu_{0}}
\int_{-\infty }^{\infty } dy'dt' 
\chi (y-y',t-t') J_{\rm J}^{z}(y',t'),
\label{veq-sw}
\end{eqnarray}
where ${c_{\rm FI}} = \sqrt {d_{\rm FI}/[(d_{\rm FI}+2\lambda_{\rm J})\epsilon _{\rm FI} \mu_{0}]}$, 
$\lambda_{\rm J} = \sqrt{\hbar/[2e\mu_{0}(d_{\rm FI} + 2\lambda_{\rm J})J_{\rm c}]}$, and 
$\Gamma_{\rm FI} = (\epsilon_{\rm FI} R_{\rm FI})^{-1}$ 
are the effective velocity of light in the FI, the Josephson penetration depth, 
and the damping factor caused by quasi-particle resistivity, $R_{\rm FI}$, in the FI, respectively. 

We look for the solution of Eq.$~$(\ref{veq-sw}) in the form 
\begin{equation}
V(y,t) = V_{0} + v(y,t)
\label{vlt},  
\end{equation}
where $V_{0}$ and $v(y,t)$ are the dc bias voltage and ac voltage induced by the ac Josephson current, respectively. 
In this case, the phase difference, $\theta(y,t)$, between two S's is given by 
\begin{equation}
\theta(y,t) = \omega_{\rm J}t - k_{\rm H}y + \theta_{1}(y,t)
\label{theta},
\end{equation}
%
%
where $\omega_{\rm J} = (2e/\hbar)V_{0}$ is the Josephson frequency, 
$k_{\rm H} =  2 \pi \mu_{0} d_{\rm FI}H_{\rm ex}/\Phi_{0}$ depends on the external magnetic field, $H_{\rm ex}$, 
and $\Phi_{0}$ is the magnetic flux quantum. 
$\theta_{1}(y,t)$ is related to $v(y,t)$ by the equation, 
\begin{equation}
v(y,t) = \frac{\hbar}{2e} \frac{\partial \theta_{1}(y,t)}{\partial t}
\label{v&theta1}.
\end{equation}
Substituting Eq.$~$(\ref{vlt}) and Eq.$~$(\ref{v&theta1}) into Eq.$~$(\ref{veq-sw}), we obtain the equation for $\theta_1(y,t)$ as follows:
\begin{eqnarray}
\frac{{\partial ^2 \theta_{1} \left( {y,t} \right)}}{{\partial y^2 }} &=& 
\frac{1}{{c_{\rm FI}^2 }}
\left[ {
{\frac{{\partial ^2 \theta_{1} \left( {y,t} \right)}}{{\partial t^2 }}}
+\frac{d_{\rm FI}}{d_{\rm FI}+2\lambda _{\rm L}} \frac{1}{\mu_{0}} 
\int_{-\infty }^{\infty } dy'dt' 
\chi (y-y',t-t') \frac{{\partial ^2 \theta_{1} \left( {y',t'} \right)}}{{\partial t^2 }} 
} \right. \nonumber \\
&+& 
\left. {
\Gamma_{\rm FI} \frac{\partial \theta_{1}(y,t)}{\partial t}
+ \Gamma_{\rm FI} \frac{d_{ \rm FI } }{d_{ \rm FI }+2\lambda _{\rm L}} \frac{1}{\mu_{0}}
\int_{-\infty }^{\infty } dy'dt' 
\chi (y-y',t-t') \frac{{\partial \theta_{1} \left( {y',t'} \right)}}{{\partial t }} 
 } \right] \nonumber \\
&+&
\frac{1}{\lambda_{\rm J}^{2}J_{\rm c}} J_{\rm J}^{z}(y,t)
+ \frac{1}{\lambda_{\rm J}^{2}J_{\rm c}} \frac{d_{\rm FI}}{d_{\rm FI}+2\lambda _{\rm L}} \frac{1}{\mu_{0}}
\int_{-\infty }^{\infty } dy'dt' 
\chi (y-y',t-t') J_{\rm J}^{z}(y',t')
\label{sgeq-sw}. 
\end{eqnarray}
We expand $\theta_{1}(y,t)$ in terms of the normal modes of the electromagnetic field 
generated by the ac Josephson current, 
\begin{equation}
\theta_{1}(y,t) = {\rm Im}
\left[
\sum_{n=1}^{\infty}
g_{n} e^{i \omega_{\rm J} t} \cos\left( k_{n} y \right)
\right]
\label{sol-theta1}, 
\end{equation}
where $g_{n}$ is a complex number and $k_{n}=n \pi /L$. 
This equation of $\theta_{1}(y,t)$ satisfies 
$[\partial \theta_{1}/\partial y]_{y=0} = [\partial \theta_{1}/\partial y]_{y=L} = 0$, 
which corresponds to the open-ended boundary condition for $v(y,t)$. 
We consider $\theta_{1}(y,t)$ to be a small perturbation and 
solve Eq.$~$(\ref{sol-theta1}) by taking $J_{\rm J}^{z} (y,t)$ to be $J_{\rm c}\sin(\omega_{\rm J}t - k_{\rm H}y)$. 
Substituting Eq.$~$(\ref{sol-theta1}) into Eq.$~$(\ref{sgeq-sw}), 
$g_{n}$ becomes (see Appendix B ) 
\begin{eqnarray}
g_{n} &=& -\frac{c_{\rm FI}^{2}}{\lambda_{\rm J}} 
		\mu (k_{\rm H}, -\omega_{\rm J})
		\frac{B_{n} - i C_{n}}{\omega_{n}^{2} - \mu(k_{n},-\omega_{\rm J}) \omega_{\rm J}^{2} +i \Gamma_{\rm FI} \mu(k_{n},-\omega_{\rm J}) 
\label{gn}}, \\
B_{n} &=& \frac{2}{L} \int_{0}^{L} dy \cos \left(k_{n}y \right) \cos\left(k_{\rm H}y\right)
\label{bn}, \\
C_{n} &=& \frac{2}{L} \int_{0}^{L} dy \cos \left(k_{n}y \right) \sin\left(k_{\rm H}y\right)
\label{cn}, \\
\mu(q,-\omega_{\rm J}) &=& 
1 +  \chi_{x} (q,-\omega_{\rm J})d_{\rm FI}/[(d_{\rm FI} + 2 \lambda_{\rm J})\mu_{0}]
\label{mu}, 
\end{eqnarray}
where $\omega_{n} = (c_{\rm FI} \pi/L)n$, and $q$ means $k_{\rm H}$ or $k_{n}$. 
In the linearized LLG equation, the magnetic susceptibility in the FI is given by (see Appendix C) 
\begin{eqnarray}
\chi_{x} \left(q, \omega_{\rm J} \right) = 
		\gamma M_{z}
		\frac{\Omega_{\rm S} + i \alpha \omega_{\rm J}}
		{\Omega_{\rm S}^{2} - (1+\alpha^{2})\omega_{\rm J}^{2} + i2\alpha \Omega_{\rm S} \omega_{\rm J}}
\label{chi-f}. 
\end{eqnarray}
Here, $\Omega_{\rm S}$ is spin wave frequency whose dispersion relation is given by 
\begin{eqnarray}
\Omega_{\rm S} &=& \Omega_{\rm B} + \frac{\eta}{\hbar} q^{2}
\label{omega-sw}, 
\end{eqnarray}
where $\Omega_{\rm B} = \gamma (H_{\rm K} - M_{z}/\mu_{0})$. 
$H_{\rm K}$ and $\eta$ are the anisotropic field and the stiffness of spin waves in the FI, respectively. 

Next, we calculate the dc Josephson current coupled with spin waves as a function of the dc voltage 
and of the external magnetic field. 
The function, $\sin(\omega_{\rm J}t - k_{\rm H}y + \theta_{1}(y,t))$, is expanded with respect to $\theta_{1}(y,t)$ 
and the dc Josephson current is given by 
\begin{equation}
J_{\rm dc} \approx  \mathop {\lim }\limits_{T \to \infty } 
		\frac{1}{T} \int_{0}^{T} dt
		\frac{1}{L} \int_{0}^{L} dy
		J_{\rm c} \cos (\omega_{\rm J} t -k_{\rm H}y) \theta_{1}(y,t)
\label{def-dc-jc}. 
\end{equation}
Introducing Eqs.$~$(\ref{sol-theta1}) and $~$(\ref{gn}) into Eq.$~$(\ref{def-dc-jc}), 
the analytic formula of the dc Josephson current is obtained as, 
\begin{eqnarray}
J_{\rm dc} &=& 
		\frac{J_{\rm c} c_{\rm FI}^{2}}{4\lambda_{\rm J}^{2}}
		\sum_{n=1}^{\infty}
		\Psi_{n} F_{n}^{2} (\phi)
\label{idc-sfs}, \\
\Psi_{n} &=& 
		{\rm Re}
		\left[
		\mu \left(k_{\rm H}, \omega_{\rm J} \right) X 
		\right]
\label{psi-sfs}, \\
X &=& \frac{1}
		{
		\omega_{n}^{2} - \mu'(k_{n},\omega_{\rm J})\omega_{\rm J}^{2} + \mu''(k_{n},\omega_{\rm J})\Gamma_{\rm FI}\omega_{\rm J}
		+i
		\left[
		\mu'(k_{n},\omega_{\rm J}) \Gamma_{\rm FI} \omega_{\rm J} + \mu''(k_{n},\omega_{\rm J}) \omega_{\rm J}^{2}
		\right]
		}
\label{x-sfs}, \\
\mu(q,\omega_{\rm J}) &=& \mu'(q,\omega_{\rm J}) + i \mu''(q,\omega_{\rm J})
\label{permeability}, \\
F_{n}^{2}(\phi) &=& 
		\left[
		\frac{2 \phi}{\phi+n/2}
		\frac{\sin\left(\pi \phi - n \pi/2 \right)}{\pi \phi - n \pi/2}
		\right]
\label{fn}, 
\end{eqnarray}
where $\phi$ is equal to $\Phi/\Phi_{0}$ and $\Phi = \mu_{0}H_{\rm ex}d_{\rm FI}L$. 
$\mu'(q, \omega_{\rm J}) = {\rm Re}[\mu(q, \omega_{\rm J})]$ and 
$\mu''(q, \omega_{\rm J}) = {\rm Im}[\mu(q, \omega_{\rm J})]$ 
(See Eqs.~(\ref{mu}) and (\ref{chi-f})). 

\begin{figure}[t]
\begin{center}
\vspace{10mm}
\includegraphics[width=9cm]{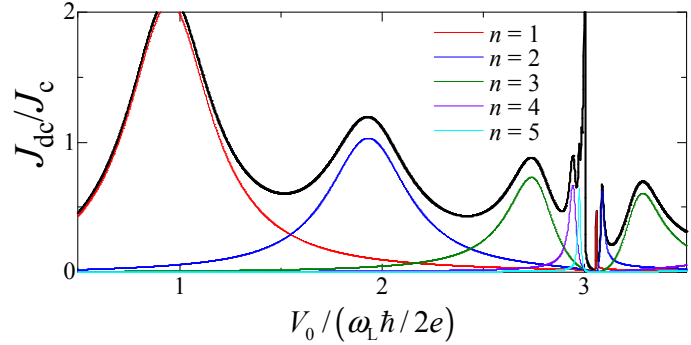}
\caption{ (Color online)
Dc Josephson current density, $J_{\rm dc}$, as a function of dc voltage, $V_{0}$, in a S/FI/S junction. 
The solid line is the total dc Josephson current. 
Red, blue, green, purple, and light blue lines are the dc Josephson current of each mode number, $n$, of electromagnetic field. 
The applied dc magnetic field determines $n$ via Eq.$~$(\ref{fn}). 
}
\label{idc-wj-sfs}
\end{center}
\end{figure}
\begin{figure}[t]
\begin{center}
\vspace{20mm}
\includegraphics[width=8cm]{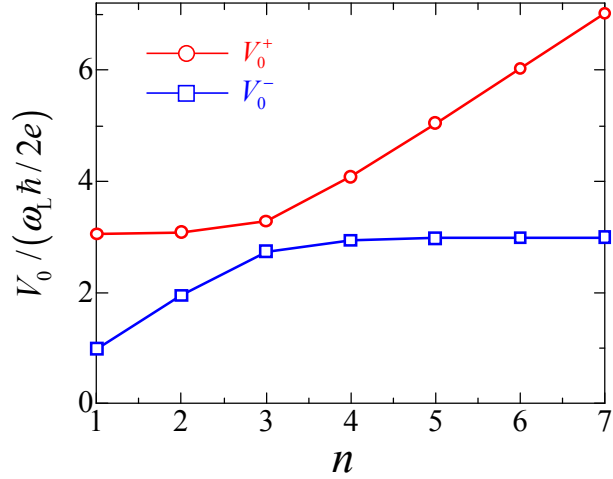}
\caption{ (Color online) 
Dc voltage, $V_{\rm 0}$, as a function of mode number, $n$, of electromagnetic field in S/FI/S junction. 
}
\label{wj-n-sfs}
\end{center}
\end{figure}

\begin{figure}[t]
\begin{center}
\vspace{20mm}
\includegraphics[width=14cm]{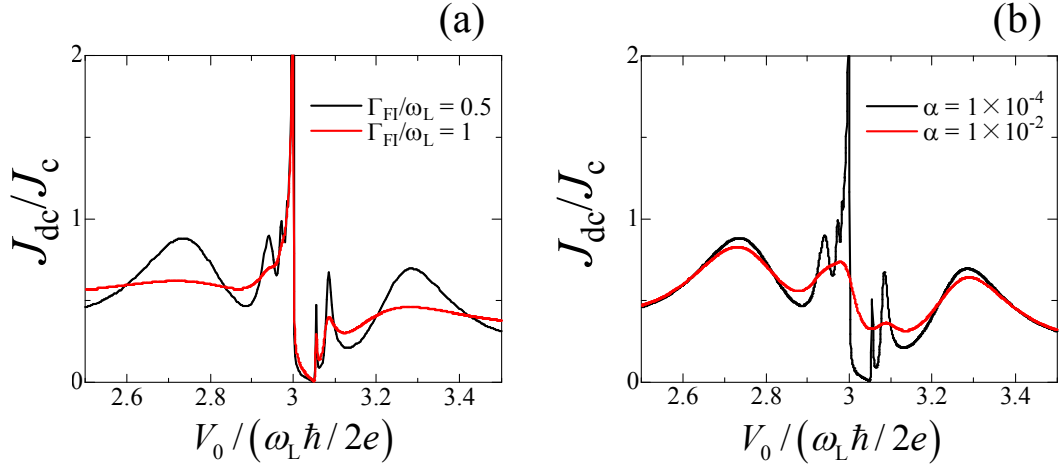}
\caption{ (Color online) 
(a) Dc Josephson current density, $J_{\rm dc}$, as a function of dc voltage, $V_{0}$, 
by changing $\Gamma_{\rm FI}/\omega_{\rm L}$ in a S/FI/S junction.
(b) $J_{\rm dc}$ as a function of $V_{0}$, by changing $\alpha$ in a S/FI/S junction.
Where parameters used by numerical calculation are $\lambda_{\rm J}/L = 1$, $\Omega_{\rm B}/\omega_{\rm L}=3$, 
$\gamma M_{z}/(\mu_{0} \omega_{\rm L}) = 1$, $d_{\rm FI}/(d_{\rm FI}+2\lambda_{\rm L}) = 0.1$.
}
\label{idc-g-a-sfs}
\end{center}
\end{figure}

\section{Results and discussion}
In this section, we examine the numerical solution for Eq.$~$(\ref{idc-sfs}). 
Figure \ref{idc-wj-sfs} shows the dc Josephson current density induced by 
the Fiske resonance as a function of the dc voltage for $\lambda_{\rm J}/L = 1$, 
$\Omega_{\rm B}/\omega_{\rm L}=3$, $\eta/(\hbar \omega_{\rm L})=3\times 10^{-16} {\rm m^{2}}$, 
$\gamma M_{z}/(\mu_{0} \omega_{\rm L}) = 1$\cite{stancil}, $\alpha=1\times10^{-4}$\cite{kajiwara}, $d_{\rm FI}/(d_{\rm FI}+2\lambda_{\rm L}) = 0.1$, 
$\Gamma_{\rm FI}/\omega_{\rm L}=3\times 10^{-1}$\cite{lisitskiy}, and $\omega_{\rm L} = c_{\rm FI} \pi/L$. 
$\phi$ is fixed as $n/2$ in the $F_{n}(\phi)$ function. 
In the solid line of Fig.$~$\ref{idc-wj-sfs}, the normalized dc Josephson current density 
$J_{\rm dc}$ in Eq.~(\ref{idc-sfs}) is shown as a function of normalized dc voltage\cite{dis2}. 
Red, blue, green, purple, and light blue lines are the dc Josephson current 
of each mode number of electromagnetic field, $n$. 
In this Figure, the resonant behavior of $J_{\rm dc}$ is due to the Fiske resonance in the S/FI/S junction. 
However, at $V_{0}/(\omega_{\rm L} \hbar/2e) \approx 3$, it is found that additional structures of $J_{\rm dc}$ appear. 
From Fig.~\ref{idc-wj-sfs}, it is found that two resonant peaks appear for each $n$. 
The appearance of the two resonant peaks in the S/FI/S junction are very different from 
conventional Josephson junctions, in which a single resonant peak appears for the each mode of EM field\cite{fiske, eck, coon, kulik, barone}. 
Large peak around $V_{0}/(\omega_{\rm L} \hbar/2e) = 3$ is due to the summation of $n > 8$ 
because dc Josephson currents of contribution from large $n$ appear around $V_{0}/(\omega_{\rm L} \hbar/2e) = 3$ in a manner to be described. 

To elucidate the origin of the resonant structures in the S/FI/S junction, we analyze Eqs.$~$(\ref{idc-sfs}) and $~$(\ref{psi-sfs}). 
When the denominator of $\Psi_n$ in Eq.$~$(\ref{psi-sfs}) is minimum with respect to 
$ \omega_{\rm J}$, $\Psi_n$ takes a maximum, 
so that the dc Josephson current shows the resonant behavior as shown in Fig.$~$\ref{idc-wj-sfs}. 
The dc voltage at which the resonance occurs is determined 
by neglecting the damping term in Eq.$~$(\ref{psi-sfs}) as $\alpha=\Gamma_{\rm FI}=0$. 
Setting the denominator of $\Psi_n$ to be zero, the voltage is given by 
\begin{eqnarray}
V_{0}^{\pm } = \frac{ \hbar }{ 2e } 
		\sqrt{ \frac{ 1 }{ 2 }
		\left[
		\omega_{n}^{2} + \Omega_{\rm S}^{2} + \frac{d_{\rm FI}}{d_{\rm FI} +
		2\lambda_{\rm L}} \frac{\gamma M_{z} \Omega_{\rm S}}{\mu_{0}} 
		\pm 
		\sqrt{ \left( \omega_{n}^{2} + \Omega_{\rm S}^{2} +
		\frac{d_{\rm FI}}{d_{\rm FI} + 2\lambda_{\rm L}} \frac{\gamma M_{z} \Omega_{\rm S}}{\mu_{0}} \right)^{2} 
		-4\omega_{n}^{2} \Omega_{\rm S}^{2} } 
		\right]
		}
\label{wj-n1}, \nonumber \\
\end{eqnarray}
where $\omega_{n}$ is the frequency of the EM field in the FI, and 
$\Omega_{\rm S}$ and $\gamma M_{z} \Omega_{\rm S}/\mu_{0}$ are 
the frequency of spin-waves and the real part of the magnetic susceptibility with $\alpha=0$ in the FI. 
We have two dc voltages, $V_{0}^{+}$ and $V_{0}^{-}$, at which the Fiske resonance occurs for each $n$. 
From the analytic formula in Eq.$~$(\ref{wj-n1}), 
it is found that two dispersions result from the coupling between the EM field and spin waves in the FI. 
Figure$~$\ref{wj-n-sfs} shows a $V_{0}$-$n$ curves obtained by Eq.$~$(\ref{wj-n1}). 
The vertical axis is the dc voltage normalized by $\omega_{\rm L} \hbar/2e$ and 
the horizontal axis is the mode number of EM field. 
In Fig.$~$\ref{wj-n-sfs}, $V_{0}^{+}$ and $V_{0}^{-}$ are shown by open circles and open squares, respectively. 
For $n<3$, $V_{0}^{+}$ is nearly constant as a function of $n$, whereas $V_{0}^{-}$ is linear with $n$. 
In $V_{0}^{+}$ for $n<3$, the voltage is nearly equal to $\Omega_{\rm S}\hbar/2e$, 
which relates to the spin wave energy in Fig.$~$\ref{wj-n-sfs}. 
For $n\geq 3$, $V_{0}^{+}$ increases as a function of $n$, whereas $V_{0}^{-}$ becomes flat with increasing $n$. 
For $n\geq 3$, $V_{0}^{-}$ is nearly equal to $\Omega_{\rm S}\hbar/2e$ in Fig.$~$\ref{wj-n-sfs}. 
Therefore, it is found that the flat behavior of the voltage comes from the spin-wave excitation in the FI. 
The spin-wave excitation in the FI is induced by the EM field generated by the ac Josephson current 
and the effect of the spin-wave excitation is reflected in the Fiske resonance in the S/FI/S junction. 

Here, we examine the $\Gamma_{\rm FI}$- and $\alpha$-dependence of $J_{\rm dc}$. 
Figure~\ref{idc-g-a-sfs} (a) shows $J_{\rm dc}$ as a function of $V_{\rm 0}$ by changing $\Gamma_{\rm FI}/\omega_{\rm L}$. 
From this figure, it is found that the Fiske resonance without very sharp peaks around $V_{0}/(\omega_{\rm L}\hbar/2e)=3$ 
exhibits strong damping by increasing $\Gamma_{\rm FI}/\omega_{\rm L}$. 
On the other hand, very sharp peaks around $V_{0}/(\omega_{\rm L}\hbar/2e)=3$ almost  does not depends on $\Gamma_{\rm FI}/\omega_{\rm L}$ 
unlike another resonant peaks. 
Next, we focus on sharp peaks around $V_{0}/(\omega_{\rm L}\hbar/2e)=3$. 
Figure~\ref{idc-g-a-sfs} (b) shows $J_{\rm dc}$ as a function of $V_{\rm 0}$ by changing $\alpha$. 
These peaks around $V_{0}/(\omega_{\rm L}\hbar/2e)=3$ in Fig.~\ref{idc-g-a-sfs} (b) strongly depend on $\alpha$ 
because these resonant peaks mainly comes from spin-waves. 
From Fig.~\ref{idc-g-a-sfs}, we can easily obtain the Fiske resonance coupled with spin-waves in the S/FI/S junction 
due to the small $\Gamma_{\rm FI}$ and $\alpha$. 

The effect of spin-waves having a finite wave number $q$ is neglected 
in the Fiske resonance because of the following reason: 
In Eq.~(\ref{omega-sw}), the first term $ \Omega_{\rm B}$ is caused by the anisotropic and 
demagnetizing fields and finite wave number $q$ is given by $n \pi/L$. 
In a conventional FI, $\hbar \Omega_{\rm B}$ is about tens of $\mu$eV\cite{hillebrands}. 
On the other hand, $\eta q^{2}$ is of the order peV due to the small stiffness of spin-waves\cite{pajda-prb} 
when the width ($L$) of the junction is a few mm. 
 
\begin{figure}[t]
\begin{center}
\vspace{10mm}
\includegraphics[width=8cm]{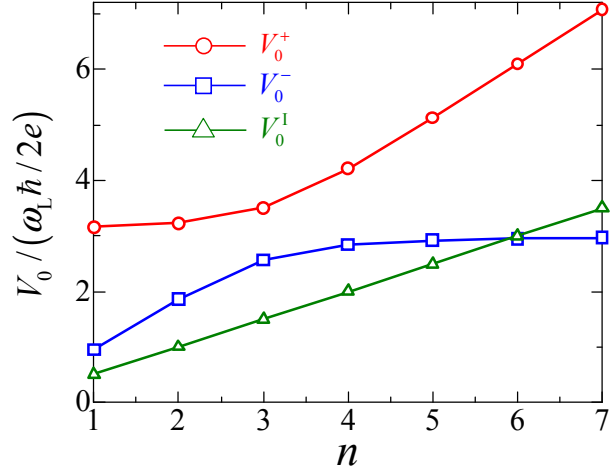}
\caption{ (Color online) 
Dc voltage, $V_{0}$, as a function of mode number, $n$, of electromagnetic field in S/I/FI/S junction. 
We take parameters as $\Omega_{\rm B}/\omega_{\rm L} = 3$, $c_{\rm I} /c_{\rm FI}= 0.5$, $\gamma M_{z}/(\mu_{0} \omega_{\rm L}) = 1$
and $d_{\rm FI}/(d_{\rm FI} + \lambda_{\rm L}) = 0.2$. 
}
\label{wj-n-sifs}
\end{center}
\end{figure}
Next, we consider the Fiske resonance in the S/I/FI/S junction. 
The details of the calculation are given by Appendix D. 
To analyze the origin of the Fiske resonance accompanied by the spin-wave excitation in the S/I/FI/S junction, 
we examine Eqs.$~$(\ref{idc-i2}) and$~$(\ref{psi-f2}). 
The condition of the resonance is given by the minimum in the denominator. 
To find out the voltage at which the Fiske resonance occurs, we neglect the damping term in the denominator of Eqs.$~$(\ref{idc-i2}) and $~$(\ref{psi-f2}). 
As a result, the voltages at the Fiske resonance are given by 
\begin{eqnarray}
V_{0}^{\rm I} &=& 
		\frac{\hbar}{2e}\omega_{n}^{\rm I}
\label{wj-i2}, \\
V_{0}^{\pm } &=& \frac{ \hbar }{ 2e }
		\sqrt{  \frac{ 1 }{ 2 }
		\left[
		\omega_{n}^{2} + \Omega_{\rm S}^{2} + \frac{d_{\rm FI}}{d_{\rm FI} +
		\lambda_{\rm L}} \frac{\gamma M_{z} \Omega_{\rm S}}{\mu_{0}} 
		\pm 
		\sqrt{ \left[ \omega_{n}^{2} + \Omega_{\rm S}^{2} +
		\frac{d_{\rm FI}}{d_{\rm FI} + \lambda_{\rm L}} \frac{\gamma M_{z} \Omega_{\rm S}}{\mu_{0}} \right]^{2} 
		-4\omega_{n}^{2} \Omega_{\rm S}^{2} }
		\right]
		}
\label{wj-f2}, \nonumber \\
\end{eqnarray}
indicating that there are three dc voltages resulting in the Fiske resonance for each $n$ in the S/I/FI/S junction. 
Figure$~$\ref{wj-n-sifs} shows the $V_{0}$-$n$ characteristic obtained by Eqs.$~$(\ref{wj-i2}) and$~$(\ref{wj-f2}). 
The vertical axis is the dc voltage normalized by $\omega_{\rm L} \hbar/2e$ and 
the horizontal axis is the mode number of EM field. 
In Fig.$~$\ref{wj-n-sifs}, $V_{0}^{\rm I}$, $V_{0}^{+}$ and $V_{0}^{-}$ are shown by 
open triangles, open circles and open squares, respectively. 
$V_{0}^{\rm I}$ linearly increases as a function of $n$ as shown in Fig.$~$\ref{wj-n-sifs} and 
comes from the resonance between the EM field in the I and the ac Josephson current. 
For $n<3$, $V_{0}^{+}$ is nearly flat as a function of $n$, whereas $V_{0}^{-}$ shows linear behavior with $n$. 
For $n\geq 3$, $V_{0}^{+}$ increases as a function of $n$, whereas $V_{0}^{-}$ becomes flat with increasing $n$. 
The voltage of flat region as a function of $n$ is nearly equal to $\Omega_{\rm S}\hbar/2e$ in Fig.$~$\ref{wj-n-sifs}. 

In this paper , we assumed that the magnetization of FI is a single domain structure. 
Usually, magnetization structure in the FI possesses a complicated domain structure. 
The ferromagnet with multidomain structure has multiple magnetic resonance modes different from the single domain system\cite{ebels}. 
Therefore, 
we can expect several additional Fiske resonance peaks in the $I$-$V$ curve arising from the domain structure in the FI. 
Those domain structure and their spin dynamics, which depend on our choice of material, have rich variety and wide spectrum. 
However, it is difficult to include the multidomain structure of FI in the present theory. 
The Fiske resonance in the ferromagnetic Josephson junction with the magnetic domains is beyond the scope of the present paper, 
and will be studied in another one. 

\section{Summary}
We have theoretically studied the coupling of Josephson-phase and spin-waves in the S/FI/S junction. 
The dc Josephson current induced by the Fiske resonance is calculated by 
combining the Maxwell and the Landau-Lifshitz-Gilbert (LLG) equations. 
The dc Josephson current shows the resonant behavior as a function of the applied dc voltage. 
We derived the analytic formula of the resonant condition in the Fiske resonance in the S/FI/S junction. 
Two resonant peaks appear in the $I$-$V$ curve for each mode number of EM field. 
We found that the two resonant peaks are generated by the coupling between spin-waves and the EM field in the FI. 
We have also studied the Fiske resonance in the S/I/FI/S junction 
and found that additional resonant structures appear due to the coupling between the ac Josephson current and the EM field in the I layer. 

\begin{acknowledgments}
The authors thank M. Aprili and T. Koyama for valuable discussions and comments. 
This work is supported by Grant in Aid for Scientific Research and  
the next Generation Supercomputer Project from MEXT. 
\end{acknowledgments}

\newpage
\appendix

\section{Derivation of Eq.$~$(\ref{veq-sw})}
We integrate Eq.$~$(\ref{faraday}) with a narrow stripe $A$ with infinitesimal width ($dy$) in the $yz$-plane, 
\begin{eqnarray}
\int_{A} dS \frac{\partial }{\partial y} E_{z}(y,t) = 
-\int_{A} dS \frac{\partial }{\partial t} \left[\mu _{0} H_{x}(y,z,t) +M_{x}(y,t) \right],
\label{s-int1}
\end{eqnarray}
where $E_{z}(y,t)$ is confined to the ferromagnetic layer due to vanishing $E_{z}(y,t)$ in the S. 
Integrating Eq.$~$(\ref{s-int1}) with respect to $y$ and 
introducing the London penetration depth $\lambda _{\rm L}$ defined by
\begin{equation}
\lambda_{\rm L} = 
\frac{1}{H_{x}(y,t)} \int_{\pm d_{\rm FI}/2}^{\pm \infty } dz H_{x}(y,z,t), \,\,\,\,\,\,\,\,\,\,\ d_{\rm FI} \ll \lambda_{\rm L},
\label{london-pene}
\end{equation}
Eq.$~$(\ref{s-int1}) becomes 
\begin{equation}
-d_{\rm FI}\frac{\partial }{\partial y} E_{z}(y,t) = 
\mu _{0} \frac{\partial }{\partial t} H_{x}(y,t) \left(d_{\rm FI}+2\lambda _{\rm L} \right) 
+ d_{\rm FI} \frac{\partial }{\partial t}M_{x}(y,t). 
\label{int-faraday}
\end{equation}
In the same way, we integrate Eq.$~$(\ref{ampere-sc}) over the cross-section area $S'$ of the junction in the $xz$-plane and obtain 
\begin{equation}
H_{x}(y,t) = \frac{1}{W} I_{\rm M}(y,t),\,\,\,\,\,\,\,\,\,\,\ I_{\rm M}(y,t) \equiv \int_{S'} dS J_{\rm M}^{y}(y,t),
\label{int-ampere-sc}
\end{equation}
where $W$ is the width along the $x$-axis of the junction and 
$I_{\rm M}(y,t)$ is the current at position $y$ in the superconducting electrode. 
Substituting Eq.$~$(\ref{int-ampere-sc}) into Eq.$~$(\ref{int-faraday}) 
and Eq.$~$(\ref{mx}), 
we obtain the partial differential equation 
\begin{equation}
\frac{\partial }{\partial y} V(y,t) = 
- \frac{\mu _{0}}{ W } (d_{\rm FI}+2\lambda _{\rm L}) \frac{\partial }{\partial t} I_{\rm M}(y,t)
+ \frac{d_{\rm FI}}{W} \frac{ \partial }{ \partial t } \int_{-\infty }^{\infty } dy'dt' \chi(y-y',t-t') I_{\rm M}(y',t'), 
\label{dvdy} 
\end{equation}
where $V(y,t) \equiv d_{\rm F}E_{z}(y,t)$ is the voltage across the FI. 
The Ampere's law in the FI is 
\begin{equation}
\frac{\partial }{\partial y} I_{\rm M}(y,t) = 
-\frac{\epsilon_{\rm FI} }{d_{\rm FI}}W \frac{\partial }{\partial t} V(y,t) 
- WJ_{\rm J}^{z}(y,t) - WJ_{\rm Q}^{z}(y,t)
\label{m-ampere-fm}
\end{equation}
Differentiating partially Eq.$~$(\ref{dvdy}) with respect to $y$, Eq.$~$(\ref{dvdy}) becomes 
\begin{equation}
\frac{ \partial^{2} }{ \partial y^{2} } V(y,t) = 
- L' \frac{ \partial }{ \partial t } \frac{ \partial }{ \partial y } I_{ \rm M }(y,t)
- L' \frac{ d_{\rm FI} }{ d_{\rm FI}+2\lambda_{\rm L} } \frac{1}{ \mu_{0} }\frac{ \partial }{ \partial t } F(y,t). 
\label{d2vdy2} 
\end{equation}
where 
\begin{equation}
F(y,t) = 
\frac{ \partial }{ \partial y } \int_{ -\infty }^{ \infty } dy'dt' \chi(y-y',t-t') I_{\rm M}(y',t'). 
\label{f-function} 
\end{equation}
To obtain the voltage equation coupled with spin waves in the FI, 
we transform the function $F(y,t)$. 
When we execute the Fourier transformation on $\chi(y-y',t-t')$ and $I_{\rm M}(y',t')$ referring to $y'$, 
$ F(y,t) $ is given by 
\begin{eqnarray}
F(y,t) &=&
\frac{1}{2 \pi} \int_{ -\infty }^{ \infty } dqdt' 
e^{ iqy } \chi(q,t-t') iq I_{\rm M}(q,t').
\label{fyt} 
\end{eqnarray}
Making use of the Fourier transformation of $y$ in Eq.$~$(\ref{m-ampere-fm}), 
\begin{eqnarray}
iq I_{\rm M}(q,t) = -C' \frac{\partial }{\partial t } V(q,t)-WJ_{\rm J}^{ z }(q,t) - WJ_{\rm Q}^{z}(q,t), 
\label{ft-m-ampere-fm}
\end{eqnarray}
and substituting Eq.$~$(\ref{ft-m-ampere-fm}) into Eq.$~$(\ref{fyt}), $F(y,t)$ is given by
\begin{eqnarray}
F(y,t) &=& 
-\int_{-\infty}^{\infty} dy'dt' \chi(y-y',t-t')
\left[
 C' \frac{\partial }{\partial t' } V(y',t') + WJ_{\rm J}^{ z }(y',t') 
+ WJ_{\rm Q}^{z}(y',t')
\right]
\label{fyt2}, 
\end{eqnarray}
Where $C' = \epsilon _{\rm FI}W/d_{\rm FI}$. 
Substituting Eq.$~$(\ref{fyt2}) into Eq.$~$(\ref{d2vdy2}) and 
using the relation $V(y,t) = d_{\rm FI} E_{z}(y,t)$, we have the partial differential equation 
\begin{eqnarray}
\frac{{\partial ^2 V \left( {y,t} \right)}}{{\partial y^2 }} &=& 
\frac{1}{{ c_{\rm FI}^2 }}
\left[ {
{\frac{{\partial ^2 V \left( {y,t} \right)}}{{\partial t^2 }}}
+\frac{d_{\rm FI}}{d_{\rm FI}+2\lambda _{\rm L}} \frac{1}{\mu_{0}} 
\frac{\partial }{\partial t}
\int_{-\infty }^{\infty } dy'dt' 
\chi (y-y',t-t') \frac{{\partial V \left( {y',t'} \right)}}{{\partial t' }}
 } \right. \nonumber \\
&+&
\left. {
\Gamma_{\rm FI} \frac{ \partial V(y,t) }{ \partial t }
+ \Gamma_{\rm FI} \frac{d_{\rm FI}}{d_{\rm FI}+2\lambda _{\rm L}} \frac{1}{\mu_{0}}
\int_{-\infty }^{\infty } dy'dt' 
\chi (y-y',t-t') V(y',t') 
 } \right ] \nonumber \\
&+&
\frac{ \hbar }{ 2e \lambda_{\rm J}^{2}J_{\rm c}} J_{\rm J}^{z}(y,t)
+ \frac{ \hbar }{ 2e \lambda_{\rm J}^{2}J_{\rm c}} \frac{d_{\rm FI}}{d_{\rm FI}+2\lambda _{\rm L}} \frac{1}{\mu_{0}}
\int_{-\infty }^{\infty } dy'dt' 
\chi (y-y',t-t') J_{\rm J}^{z}(y',t') 
\label{voltage-eq},
\end{eqnarray}

\section{Calculation of factor $g_{n}$}
To obtain the complex coefficient $g_{n}$, substituting Eq.$~$(\ref{sol-theta1}) and $J_{\rm J}\approx J_{\rm c} \sin(\omega_{\rm J} t -k_{\rm H}y)$ 
into Eq.$~$(\ref{sgeq-sw}), 
we obtain the equation for $g_{n}$, 
\begin{eqnarray}
\sum_{n=1}^{\infty}
		g_{n} e^{i \omega_{\rm J} t} \cos\left(k_{n}y \right)
		\left[
		-\omega_{n}^{2} + \mu(k_{n},-\omega_{\rm J}) \omega_{\rm J}^{2} - i \Gamma_{\rm FI} \mu(k_{n},-\omega_{\rm J}) \omega_{\rm J}
		\right]
 =
		\frac{c_{\rm FI}^{2}}{\lambda_{\rm J}^{2}} \mu(k_{\rm H},-\omega_{\rm J}) e^{i \omega_{\rm J}t-ik_{\rm H}y}
\label{gn-ap1}, 
\end{eqnarray}
where $\omega_{n} = (c_{\rm FI} \pi/L)n$ and 
$\mu(q,-\omega_{\rm J}) = 1 +  \chi (q,-\omega_{\rm J})d_{\rm FI}/[(d_{\rm FI} + 2 \lambda_{\rm J})\mu_{0}]$ and 
$q$ denotes $k_{\rm H}$ or $k_{n}$. 
For both sides of Eq.$~$(\ref{gn-ap1}), multiplying $\cos(k_{n}y)$ and performing integration with respect to $y$ from $0$ to $L$, 
we get $g_{n}$ as follows:
\begin{eqnarray}
g_{n} &=& -\frac{c_{\rm FI}^{2}}{\lambda_{\rm J}} 
		\mu (k_{\rm H}, -\omega_{\rm J})
		\frac{B_{n} - i C_{n}}{\omega_{n}^{2} - \mu(k_{n},-\omega_{\rm J}) \omega_{\rm J}^{2} +i \Gamma_{\rm FI} \mu(k_{n},-\omega_{\rm J}) 
\label{gn-apb}}, \\
B_{n} &=& \frac{2}{L} \int_{0}^{L} dy \cos \left(k_{n}y \right) \cos\left(k_{\rm H}y\right), \\
C_{n} &=& \frac{2}{L} \int_{0}^{L} dy \cos \left(k_{n}y \right) \sin\left(k_{\rm H}y\right).  
\end{eqnarray}

\section{Linearized solution of LLG equation}
We outline the derivation of the magnetic susceptibility in the FI by using the LLG equation. 
We consider the situation, in which the direction of magnetization is perpendicular to the junction.
In the experimental measurement of the Fiske resonance, the dc magnetic field is smaller than several tens of gauss. 
Therefore, we can neglect the gradient of the magnetization due to the applied dc magnetic field. 
Since the precessional angle of spin in the FI is usually very small even at the magnetic resonance, 
we linearize the LLG equation as follows: 
%
\begin{eqnarray}
\frac{dM_{x}}{dt} &=&
		-\gamma M_{y}
		\left(
		H_{K} - M_{z}/\mu_{0} 
		\right)
		+
		\gamma M_{z} \Lambda \partial_{y}^{2} M_{y}
		-\alpha \frac{dM_{y}}{dt}
\label{apd-mx-eq}, \\
\frac{dM_{y}}{dt} &=&
		\gamma M_{x} 
		\left(
		H_{K} - M_{z}/\mu_{0}
		\right)
		-\gamma M_{z} h_{x} 
		-\gamma M_{z} \Lambda \partial_{y}^{2} M_{x}
		+\alpha \frac{dM_{x}}{dt}
\label{apd-my-eq}, 	
\end{eqnarray}
where $H_{K}$ and $M_{z}/\mu_{0}$ are an anisotropic field and a demagnetization field in the FI, respectively. 
We assume the solutions of Eq.$~$(\ref{apd-mx-eq}) and $~$(\ref{apd-my-eq}) of the form 
\begin{eqnarray}
M_{i} &=& \chi_{i} (q,\omega_{\rm J}) h_{x} e^{i \omega_{\rm J} t} e^{-i q y},\,\,\,\,\,\,\ i = x, y,
\label{apd-mi-sol}
\end{eqnarray}
where $\chi_{i}(q, \omega_{\rm J})$ is a magnetic susceptibility of FI 
and $h_{x} $ is the amplitude of $x$-component of ac magnetic field and $q$ is a wave number of spin wave. 
Substituting Eq.$~$(\ref{apd-mi-sol}) into Eq.$~$(\ref{apd-mx-eq}) and (\ref{apd-my-eq}), 
the solutions of the linearized LLG equation are
\begin{eqnarray}
M_{x} &=& \chi_{x} (q, \omega_{\rm J}) h_{x},
\label{apd-mx-sol} \\
\chi_{x}(q,\omega_{\rm J}) &=& 
		\gamma M_{z}
		\frac{\Omega_{\rm S} + i \alpha \omega_{\rm J}}
		{\Omega_{\rm S}^{2} - (1+\alpha^{2})\omega_{\rm J}^{2} + i2 \alpha \Omega_{\rm S} \omega_{\rm J}}, \\
M_{y} &=& \chi_{y} (q, \omega_{\rm J}) h_{x},
\label{apd-mx-sol} \\
\chi_{y}(q,\omega_{\rm J}) &=& 
		-\gamma M_{z}
		\frac{i\omega_{\rm J}}{\Omega_{\rm S}^{2} - (1+\alpha^{2})\omega_{\rm J}^{2} + i2 \alpha \Omega_{\rm S} \omega_{\rm J}}, \\
\Omega_{\rm S} &=& \Omega_{\rm B} + \frac{\eta}{\hbar} q^{2}, \,\,\,\,\,\,\,\,\, \Omega_{\rm B} = \gamma (H_{\rm K} - M_{z}/\mu_{0}),
\end{eqnarray}
where $H_{\rm K}$ is an anisotropic field in the FI and $\eta$ is a stiffness of spin waves. 

\section{Derivation of dc Josephson current in S/I/FI/S junction}
In this Appendix, we consider the Josephson junction including a nonmagnetic insulator (I) between the S and the FI, that is the S/I/FI/S junction. 

To calculate the Josephson current in the S/I/FI/S junction, 
we solve the Maxwell equation as shown in the section II to add to equations describing the I as follows:
\begin{eqnarray}
{\rm rot} \left[ E_{z}^{\rm I}(y,t) {\bm e_{z}} \right] &=& 
-\mu _{0} \frac{\partial }{\partial t} \left[ H_{x}^{\rm I}(y,z,t) \right] {\bm e}_{x},
\label{faraday-i} \\
{\rm rot} \left[ H_{x}^{\rm I}(y,t) {\bm e}_{x} \right] - 
\epsilon_{\rm I} \frac{\partial }{\partial t} \left[ E_{z}^{\rm I}(y,t) {\bm e}_{z} \right] &=&
J_{\rm J}^{z}(y,t){\bm e}_{z} + \sigma_{\rm I}E_{z}^{\rm I}(y,t){\bm e}_{z} 
\label{ampere-i},
\end{eqnarray}
where the subscript I in the above equations indicates fields in the non-magnetic insulator. 
$\epsilon_{\rm I}$ and $\sigma_{\rm I}$ are a dielectric constant and a conductivity in the I, respectively. 
$J_{\rm J}^{z}$ is a Josephson current. 
In the same manner as the calculation in the section II, 
we integrate the Faraday's law with respect to the plane parallel to the $yz$-plane and 
the Ampere's law with respect to the plane parallel to the $xz$-plane. 
As a result, Ampere's and Faraday's laws are given by
\begin{eqnarray}
-\frac{\partial E_{z}^{\rm FI} (y,t)}{\partial y}d_{\rm FI} - \frac{\partial E_{z}^{\rm I} (y,t)}{\partial y}d_{\rm I} &=&
		\mu_{0} \frac{ \partial H_{x}^{\rm FI}(y,t) }{\partial t} \left( d_{\rm FI} + \lambda_{\rm L} \right) 
		+ \frac{ \partial M_{x}(t) }{ \partial t } d_{\rm FI} \nonumber \\
		&+&\mu_{0} \frac{ \partial H_{x}^{\rm I}(y,t) }{\partial t} \left( d_{\rm I} + \lambda_{\rm L} \right)
\label{faraday-2}, \\
H_{x}^{i}(y,t) &=& \frac{1}{W} I_{\rm L}^{i}, \,\,\,\,\,\ i= {\rm FI}\,\,\ {\rm or} \,\,\ {\rm I}
\label{ampere-2}. 
\end{eqnarray}
The definition of $I_{\rm L}^{i}$ is same as that of Eq.$~$(\ref{int-ampere-sc}). 
Performing the procedure of calculation to obtain Eq.$~$(\ref{veq-sw}) for Eqs.$~$(\ref{faraday-2}) 
and (\ref{ampere-2}), 
we obtain equations describing ac voltages induced by the ac Josephson current in the I and the FI layers as, 
\begin{eqnarray}
\frac{ \partial^{2} V^{\rm I } (y,t) }{ \partial y^{ 2 } } &=& 
		\frac{1}{ c_{\rm I}^{2}} \frac{ \partial^{2} V^{\rm I}(y,t) }{ \partial^{2} t }
		+
		\frac{1}{ c_{\rm I}^{2}} \Gamma_{\rm I} \frac{\partial V^{\rm I}(y,t) }{\partial t}
		+
		\frac{\hbar }{2e} \frac{1}{\lambda_{\rm I}^{2} J_{\rm c} } \frac{ \partial J_{\rm J}(y,t) }{ \partial t }
\label{eq-v-i},  \\
c_{\rm I}^{-2} &=& \frac{d_{\rm I} + \lambda_{\rm L} }{d_{\rm I}} \epsilon_{\rm I} \mu_{0}, 
\,\,\ \lambda_{\rm I}^{-2} = \frac{2e}{\hbar} \mu_{0} \left( d_{\rm I} + \lambda_{\rm L} \right), 
\,\,\ \Gamma_{\rm I} = \frac{1}{\epsilon_{\rm I} R_{\rm I} }, \\
\frac{ \partial^{2} V^{\rm FI} (y,t) }{ \partial y^{ 2 } } &=& 
		\frac{1}{c_{\rm FI}^{2}} \frac{ \partial^{2} V^{\rm FI}(y,t) }{ \partial^{2} t }
		+
		\frac{1}{c_{\rm FI}^{2}}\frac{d_{\rm FI}}{d_{\rm FI} + \lambda_{\rm L}} 
		\frac{\partial}{\partial t} \int_{-\infty}^{\infty} dt'
		\chi(t-t') \frac{\partial V^{\rm FI}(y,t') }{\partial t } \nonumber \\
		&+&
		\frac{1}{c_{\rm FI}^{2}} \Gamma_{\rm FI} \frac{\partial V^{\rm FI}(y,t) }{\partial t}
		+\frac{\partial}{\partial t} \int_{-\infty}^{\infty} dt'
		\chi(t-t') V^{\rm FI}(y,t') \nonumber \\
		&+&
		\frac{\hbar }{2e} \frac{1}{\lambda_{\rm FI}^{2} J_{\rm c} } \frac{ \partial J_{\rm J}(y,t) }{ \partial t }
		+
		\frac{ \hbar }{ 2e } \frac{ 1 }{ \lambda_{\rm FI}^{2} J_{\rm c} } \frac{ d_{\rm FI} }{ d_{\rm FI } + \lambda_{\rm L}}
		\frac{\partial }{\partial t} \int_{-\infty}^{\infty} dt' \chi(t-t')J_{\rm J}(y,t'), 
\label{eq-v-fm} \\
c_{\rm FI}^{-2} &=& \frac{d_{\rm FI} + \lambda_{\rm L} }{d_{\rm FI}} \epsilon_{\rm FI} \mu_{0}, 
\,\,\ \lambda_{\rm FI}^{-2} = \frac{2e}{\hbar} \mu_{0} \left( d_{\rm FI} + \lambda_{\rm L} \right), 
\,\,\ \Gamma_{\rm FI} = \frac{1}{\epsilon_{\rm FI} R_{\rm FI} }, 
\end{eqnarray}
where $d_{\rm I}$ is the thickness of I. 
$\epsilon_{\rm I}$ and $R_{\rm I}$ are a dielectric constant and resistance per unit length in the I. 
Here, we adopt the iterative calculation to obtain the dc Josephson current due to the resonance between 
the electromagnetic field and Josephson current as shown in previous section. 
First, we calculate the voltage induced by the ac Josephson current  in each layers. 
Josephson current is characterized by the phase difference between two S's. 
Therefore, we assume that the Josephson current flowing layers is given by 
$J_{\rm J}(y,t) = J_{\rm c} \sin(\omega_{\rm J}t-k_{\rm H}y) $, 
where $k_{\rm H}=2 \pi \mu_{0}H_{\rm ex} (d_{\rm FI} + d_{\rm I})/\Phi_{0}$ and $\omega_{\rm J}$ is the Josephson frequency. 
For the boundary condition, we adopt the open-ended boundary condition for the reason mentioned in section II. 
In this case, the voltage expression satisfying the open-ended boundary condition is given by 
\begin{eqnarray}
V^{j}(y,t) = {\rm Im}
		\left[
		\sum_{n=1}^{\infty}
		v_{n}^{j} e^{i \omega_{\rm J} t}
		\cos\left( k_{n} y \right)
		\right], \,\,\ j = {\rm I\,\ or \,\ FI}
\label{sol-vif}, 
\end{eqnarray}
where $v_{n}^{j}$ are complex numbers and $k_{n}=n \pi/L$.  
Substituting Eq.$~$(\ref{sol-vif}) into Eqs.$~$(\ref{eq-v-i}) and (\ref{eq-v-fm}), 
we can obtain the voltage in I and F layers as follows:
\begin{eqnarray}
V^{\rm I} (y,t) &=& \frac{\hbar}{eL} \left(\frac{c_{\rm I} }{ \lambda_{\rm I} }\right)^{2}
		\sum_{n=1}^{\infty} 
		\left[
		\frac{-\Gamma_{\rm I} \omega_{\rm J}^{2} B_{n} + \omega_{\rm J} \left[ \omega_{\rm J}^{2} - (\omega_{n}^{\rm I})^{2} \right] C_{n} }
		{\left[ \omega_{\rm J}^{2} - (\omega_{n}^{\rm I})^{2} \right]^{2} +\left(\Gamma_{\rm I} \omega_{\rm J} \right)^{2} }
		\sin \left(\omega_{\rm J} t \right)
		\right. \nonumber \\
		&+&
		\left.
		\frac{\omega_{\rm J} \left[\omega_{\rm J}^{2} - (\omega_{n}^{\rm I})^{2} \right]B_{n} + \Gamma_{\rm I} \omega_{\rm J}^{2} C_{n} }
		{\left( \omega_{n}^{\rm I} - \omega_{\rm J} \right)^{2} +\left(\Gamma_{\rm I} \omega_{\rm J} \right)^{2} }
		\cos \left(\omega_{\rm J} t \right)
		\right]
		\cos (k_{n}y)
\label{vi2}, \\
V_{\rm FI} (y,t) &=& -\frac{\hbar}{eL} \left(\frac{ c_{\rm FI} }{ \lambda_{\rm I} }\right)^{2}
		\sum_{n=1}^{\infty}
		\left[ \Re[g(\omega_{\rm J}) \sin(\omega_{\rm J}) + \Im[g(\omega_{\rm J})] \cos(\omega_{\rm J} t) \right)] \cos(k_{n}y)
\label{vf2}, \\
g_{n} &=& -
		\left(
		\frac{c_{\rm FI}}{\lambda_{\rm J}} 
		\right)^{2} 
		\mu (-\omega_{\rm J})
		\frac{B_{n} - i C_{n}}{\omega_{n}^{2} - \mu(-\omega_{\rm J}) \omega_{\rm J}^{2} +i \Gamma_{\rm FI} \mu(-\omega_{\rm J}) 
\label{gn2}}, 
\end{eqnarray}
where $\omega_{n}^{\rm I}$ and $\omega_{n}$ are given by $({c}_{\rm I} \pi/L)n$ and $({c}_{\rm FI} \pi/L)n$, respectively. 
$B_{n}$ and $C_{n}$ are same expressions as Eq.$~$(\ref{bn}) and Eq.$~$(\ref{cn}), respectively. 

Next we calculate the dc Josephson current coupled with spin waves as a function of the dc voltage and of the external magnetic field. 
Since we consider $V^{\rm I (\rm FI)}(y,t)$ as a perturbation, we can expand the sine function in terms of $V^{\rm I (\rm FI)}(y,t)$. 
Within the first order term with respect to $V^{\rm I (\rm FI)}(y,t)$, the dc Josephson current is approximately given by
\begin{eqnarray}
J_{\rm dc} \approx   \mathop {\lim }\limits_{T \to \infty } 
		\frac{1}{T} \int_{0}^{T} dt
		\frac{1}{L} \int_{0}^{L} dy
		J_{\rm c} \cos \left( \omega_{\rm J}t - k_{\rm H}y \right) \frac{2 \pi}{\Phi_{0}} \int dt \left[V^{\rm I}(y,t) + V^{\rm FI}(y,t) \right]
\label{jdc2}. 
\end{eqnarray}
Substituting Eqs.$~$(\ref{vi2}) and (\ref{vf2}) into Eq.$~$(\ref{jdc2}), 
we can obtain the analytic formula of the dc Josephson current in the S/I/FI/S junction as follows: 
\begin{eqnarray}
J_{\rm dc} &=& J_{\rm dc}^{\rm I} + J_{\rm dc}^{\rm FI}, \\
J_{\rm dc}^{\rm I} &=& \frac{J_{\rm c} c_{\rm I}^{2} }{ 4\lambda_{\rm I}^{2} } 
		\sum_{n=1}^{\infty} 
		\frac{\Gamma_{\rm I} \omega_{\rm J} }{ \left[(\omega_{n}^{\rm I})^{2} - \omega_{\rm J}^{2}\right]^{2} + \left(\Gamma_{\rm I} \omega_{\rm J} \right)^{2} }
		F_{n}^{2} \left( \phi \right)
\label{idc-i2}, \\
J_{\rm dc}^{\rm FI} &=& 
		\frac{J_{\rm c} c_{\rm FI}^{2}}{4\lambda_{\rm FI}^{2}}
		\sum_{n=1}^{\infty}
		\Psi_{n}^{\rm F} F_{n}^{2} (\phi)
\label{idc-f2}, \\
\Psi_{n}^{\rm F} &=& \frac{\mu'(\omega_{\rm J}) 
		\left[
		\mu'(\omega_{\rm J}) \Gamma_{\rm FI} \omega_{\rm J} + \mu''(\omega_{\rm J}) \omega_{\rm J}^{2}
		\right]
		+ \mu''(\omega_{\rm J})
		\left[
		\omega_{n}^{2} - \mu'(\omega_{\rm J})\omega_{\rm J}^{2} + \mu''(\omega_{\rm J})\Gamma_{\rm FI}\omega_{\rm J}
		\right]}
		{\left[
		\omega_{n}^{2} - \mu'(\omega_{\rm J})\omega_{\rm J}^{2} + \mu''(\omega_{\rm J})\Gamma_{\rm FI}\omega_{\rm J}
		\right]^{2}
		+
		\left[
		\mu'(\omega_{\rm J}) \Gamma_{\rm FI} \omega_{\rm J} + \mu''(\omega_{\rm J}) \omega_{\rm J}^{2}
		\right]^{2}}	
\label{psi-f2}, \nonumber \\
\end{eqnarray}
where $\omega_{n}^{\rm I}$ and $\omega_{n}$ are given by $({c}_{\rm I} \pi/L)n$ and $({c}_{\rm FI} \pi/L)n$, respectively. 
$F_{n}(\phi)$ is same function as Eq.$~$(\ref{fn}). 
$J_{\rm JI}^{\rm dc}$ comes from the resonance between the electromagnetic field generated inside the I and the ac Josephson current. 
$J_{\rm JFI}^{\rm dc}$ originates in the resonance between the electromagnetic field generated inside the FI and the ac Josephson current. 


\begin{thebibliography}{}
\bibitem{dis2}
For $L=1$ $mm$ and $c_{\rm FI}=10^{7}$ $\rm m/\rm s$, the magnitude of dc voltage ($V_{0}$) is about tens of $\mu V$.
\end{thebibliography}
\end{document}